\documentclass[a4paper]{jpconf}

\usepackage{graphicx}
\usepackage{amsmath,amssymb}

\usepackage{citesort}

\usepackage{setspace}

\begin{document}


\title{Electron emission from a metal nano-tip by ultrashort laser pulses}

\author{G Wachter$^1$, C Lemell$^1$ and J Burgd\"orfer$^{1}$}
\address{$^1$Institute for Theoretical Physics, Vienna University of Technology, Wiedner Hauptstr. 8-10, A-1040 Vienna, Austria, EU}
\ead{georg.wachter@tuwien.ac.at}
\begin{abstract}
We theoretically investigate the interaction of near-infrared few cycle laser pulses of moderate intensity with nano-scale metal tips. Macroscopic field enhancement leads to coherent electron emission from the tip apex. Electron spectra are simulated with time-dependent density functional theory (TDDFT). We investigate the dependence of the simulated electron spectra on the choice of exchange-correlation potential and atomic core pseudo-potential.
\end{abstract}

\section{Introduction}
Recent experimental and theoretical studies have shown \cite{Schenk2010StrongField,Kruger2011Attosecond,Wachter2012Electron} that a nano-tip illuminated by near-infrared moderate intensity few cycle laser pulses affords the possibility to study the effect of a strong time-dependent field acting on a solid-state system. The quantum dynamics near the surface can be inferred from analyzing the electrons which are photo-emitted by the laser pulse. The measured electron energy spectra show clear signs of strong field physics even at low nominal laser intensities which can be traced back to the field enhancement near the apex of the tip due the so-called lightning-rod effect. A sub-wavelength sized region near the tip apex experiences an enhanced laser field (field enhancement factor $\sim 5$ for tip radius $\sim 10$ nm). This allows for the observation of strong field effects at low nominal laser intensities and also lies at the heart of the observability of coherence effects as the emitted electrons originate from a tightly confined region in both space and time. The coherence of the emission process is reflected in the electron energy spectra, their carrier-envelope phase dependence and their angular distribution. Within a semi-classical description, the observed multi-photon peaks can be traced back to the periodic emission of coherent electron wave packets equi-spaced in time given by the laser period, leading to interference modulation of the energy spectra with the period of the photon energy  \cite{Kruger2012Interaction}. Another feature typically associated with strong-field physics is found in the high-energy part of the electron spectra showing a plateau region followed by a cut-off. The latter stem from electrons undergoing a process known as ``rescattering'' in strong field physics for atoms: first, the electron tunnels out from its parent core near a maximum of the laser field. Subsequently, it is driven back towards the core as the laser field changes sign attaining energies of $\sim 3 U_{\mathrm{p}}$ with the ponderomotive potential $U_{\mathrm{p}} = F_0^2 / 4 \omega^2$ for laser amplitude $F_0$ and laser frequency $\omega$. Upon rescattering in backward direction from the potential gradient near the core, the electron gains additional energy up to a final energy of $10 \, U_{\mathrm{p}}$ determining the position of the observed cut-off \cite{Paulus1994Rescattering}. The high-energy part of electron spectra from nano-tips has been observed \cite{Kruger2011Attosecond} and its origin can unambiguously be traced to such a process \cite{Wachter2012Electron,Kruger2012Interaction}.

The plan for this paper is as follows: first, we review our quantum mechanical model for electron emission from a nano-tip based on time-dependent density functional theory (TDDFT). We show a sample simulation result where the rescattering process can be directly observed in the time-dependent electron density. We then focus on the dependence of the simulated electron spectra on two parameters entering the TDDFT description of the process, the exchange-correlation potential and the atomic core pseudo-potential used to simulate the outermost layer of atoms. Atomic units are used unless stated otherwise.

\section{Time-dependent density functional theory for electron emission from nano-tips}
We briefly review our microscopic description of electron emission based on time-dependent density functional theory (TDDFT) \cite{Liebsch1997Electronic,Burke1998Density,Maitra2002Ten,Lemell2003Electron}. Since in the experiment the tip radius ($\sim 10$ nm) is much larger than the length scale of the electronic system (Fermi wavelength $\sim0.1$ nm), translational symmetry in the surface plane is approximately conserved on an atomic scale. We hence restrict ourselves to a one-dimensional model, treating the coordinate along the surface normal $z$ as reaction coordinate along which the macroscopic field enhancement is observed. A detailed description is given in \cite{Wachter2012Electron}. In short, the time-dependent electron density $n(z,t)$ is expanded into Kohn-Sham orbitals $\psi_k(z,t)$
\begin{equation}
n(z,t)=\sum_{k=1}^{n_{\mathrm{occ}}} c_k |\psi_k(z,t)|^2 \label{eq2} \quad ,
\end{equation}
where $n_{\mathrm{occ}}$ is the number of occupied orbits up to the Fermi energy $E_{\mathrm{F}}$. We use a metal slab of $\sim$200 a.u.~width, yielding $n_{\mathrm{occ}}\sim 50$ orbitals representing the conduction band. The weight coefficients $c_k$ are derived from the projection of the three-dimensional Fermi sphere onto the tip axis \cite{Eguiluz1984Static} such that eq.\ \ref{eq2} gives the ground state density for $t\to -\infty$.  The electron density is expressed in terms of the Wigner-Seitz radius $r_\mathrm{s}=2.334$  a.u.~according to the s-electron density of tungsten giving a Fermi energy of $E_\mathrm{F}=9.2$ eV. 

The time evolution of the electron density is given by the time-dependent coupled Kohn-Sham equations
\begin{equation} \label{eq:tdks}
i \partial_t\psi_k(z,t) = \left\{-\frac{1}{2} \partial^2_z + V[n(z,t)] + V_{\mathrm{ext}}(z,t)\right\}\psi_k(z,t)\, ,
\end{equation}
where $V[n(z,t)]$ contains the ground state, electrostatic, and exchange-correlation potentials. For the ground state we use a smooth jellium effective single particle potential of the form  \cite{Burgdorfer1994Atomic} 
\begin{equation}
V_{\mathrm{jellium}} (z) = 
    \begin{cases}
     - \frac{ V_0 }{ A \exp( B z ) + 1 }  \, ,  z < z_{im}    \quad \text{,} \\
       \frac{ 1 - \exp( - b z ) } { 4 z }     \, , z \geq z_{im} 
     \end{cases}
\end{equation}
where the depth of the potential well is adjusted to $V_0 = E_{\mathrm{F}} + W $ with the work function $W$, the position of the image plane relative to the jellium edge is taken as $ z_{im} = -0.2 r_s + 1.25$, the parameter governing the transition to the asymptotically correct $1/(4 z)$ image potential is chosen as $b = k_F$, and consequently $A = 4 V_0 / b - 1 $ and $B = V_0 / ( 4 V_0 / b - 1)$. This ground state potential has the advantage that the work function can be continuously adjusted. A clean tungsten (310) surface has a work function of $W_{(310)}=4.35$ eV, which is, however, sensitive to surface adsorbates~\cite{Yamamoto1978Field} and can serve only as a first estimate. Best agreement to experiment was found for a work function of $W = 6.2$ eV \cite{Wachter2012Electron} which is also used throughout this work. Apart from the value of the work function, the exact shape of the ground state potential is of minor importance. We have tested other potentials including potentials accurately reproducing the electronic surface structure for a number of materials \cite{Chulkov1999Image} yielding similar results.

To account for rescattering, a localized potential simulating the atomic cores of the first atomic layer is added to the jellium potential (see section \ref{sec:surfpot}). It is parametrized by a screened soft-core Coulomb potential
\begin{equation} \label{eq:vatom}
V_{\mathrm{atom}}(z)=- \frac{1}{1+|z|}  e^{-|z|/\lambda_{\mathrm{TF}}}
\end{equation}
with the Thomas-Fermi screening length $\lambda_{\mathrm{TF}}\approx 1$ a.u.\ for the electron gas and is centered at the position of the first atom layer, i.e.\ one half of the effective lattice position $a \times \frac{\sqrt 3 }{4} \approx 2.6 $ a.u.~inside the jellium edge for a tungsten (310) surface.

The electrostatic potential is derived from the Poisson equation in one dimension, $ \frac{ d^2 }{dz^2 } V_{es} (z,t) = - 4 \pi n(z,t)$. For the exchange-correlation potential, we assume the local density approximation (LDA) in the Perdew-Zunger parametrization \cite{Perdew1981Selfinteraction} by default. The dependence of simulation results on this parameter is studied in section \ref{sec:vxc}. The external potential from the laser pulse is given in dipole approximation $V_{\mathrm{ext}}(z,t) = z F_{\mathrm{eff}}(t) + z F_{\mathrm{dc}}$ where the enhancement due to the lightning rod effect is already included in $F_{\mathrm{eff}}$ and $F_{\mathrm{dc}}$ \cite{Wachter2012Electron}. As in experiment, a small static extraction field $F_{\mathrm{dc}}$ is included which is adiabatically switched on before the laser pulse starts.

The Kohn-Sham equations (\ref{eq:tdks}) are integrated in real space by the Crank-Nicolson method with a constant time step of 0.05 a.u., extrapolating the time-dependent Hamiltonian at $t+\Delta t/2$ to second order in time. The simulation time is typically 120 fs ($\sim 5000$ a.u.) which is chosen large enough so that electrons of low energy have passed the detection point (see below). The size of the simulation box was set to 1425 a.u.\ (9500 grid points) including absorbing potentials near the borders to avoid unphysical reflections. Electron energy spectra are determined \cite{Pohl2000Towards} by a temporal Fourier transform of the wave functions at a detection point far from the surface ($\sim 900$ a.u.) ensuring that the laser pulse has terminated at the time of arrival of the wavepacket. More information on calculating electron spectra from TDDFT can be found in \cite{DeGiovannini2012TextitAb,Dinh2012Critical}. The calculated spectra are broadened by $0.5$ eV, corresponding to a typical  experimental spectrometer resolution. The electron energy is measured from the Fermi energy since the vacuum level is not well-defined in the presence of a static field.

\section{Sample simulation} 
\label{sec:samplesim}
\begin{figure}[h] 
\centerline{\includegraphics[width=0.9\textwidth]{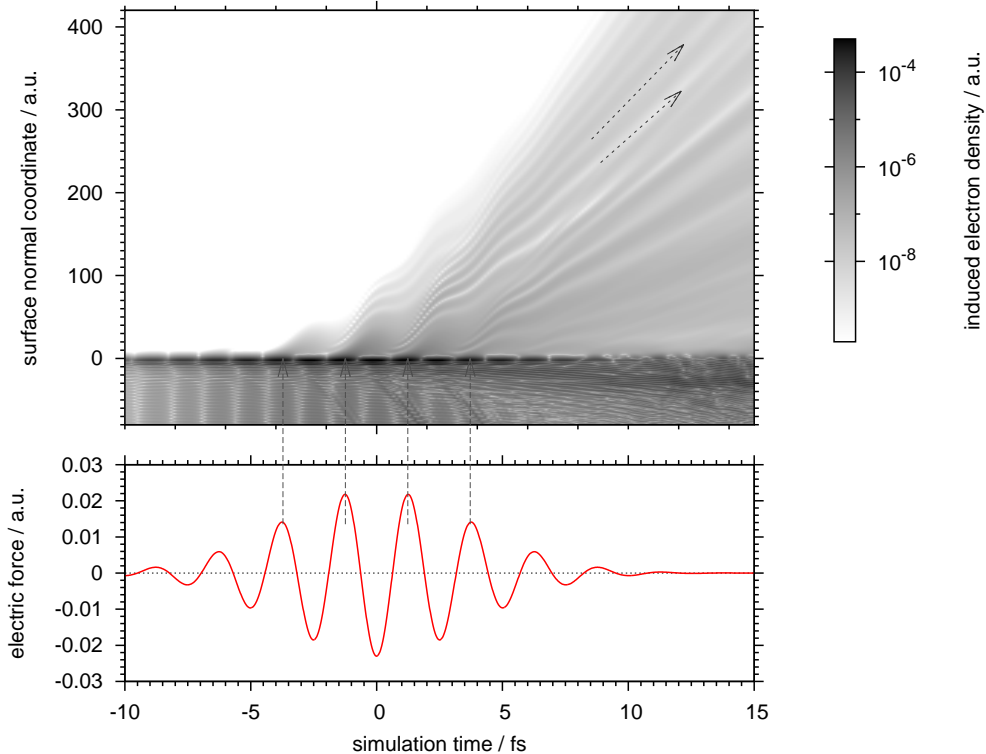}}
\caption{\label{fig:tddens}
Sample simulation. Top panel: Absolute value of the time-dependent induced density $|n(z,t)-n(z,-\infty)|$ on a logarithmic scale (for comparison, the bulk background density is $n_0 = 1.88 \times 10^{-2}$ a.u.). Bottom panel: Time-dependent external electric field (red solid line). Ionization happens mostly at the maxima of the electric field (marked by vertical arrows). Electrons are subsequently driven back to the surface and rescatter. Interferences in position space with increasing slopes (velocities) are detected as equi-spaced multi-photon peaks in the energy spectra (marked by tilted arrows).
}
\end{figure}

A typical simulation run is shown in fig.\ \ref{fig:tddens} for a laser intensity close to the experimental destruction threshold with a laser amplitude of $F_{\mathrm{eff}}= 0.023$ a.u.$= 9.77 $ GV/m ($I_{\mathrm{eff}} = 1.9 \times 10^{13}$ W/cm$^\mathrm{2}$). Other parameters are chosen in accordance with experiment \cite{Wachter2012Electron}. The pulse length is  6.4 fs (FWHM of the intensity envelope) with a photon energy of  $\hbar \omega = 0.06$ a.u. corresponding to a near-infrared wavelength of $\lambda = 760$ nm. The carrier-envelope phase is chosen as $\phi_{CEP}= +\pi$ (a $\left( -\cos \right)$-pulse for the force), and a small static extraction field of $F_{\mathrm{dc}} = 0.00019 $ a.u.~$= 0.1$ GV/m $ \approx 1 $\% $ F_{\mathrm{eff}}$ is applied.

Electron density is emitted at the field maxima from the tip ($ \lesssim 0$) into vacuum ($ \gtrsim 0$) (grey vertical arrows in fig.\ \ref{fig:tddens}). As the laser field changes sign, electrons are driven back towards the surface. They are reflected from the surface approximately at the zero crossings of the electric field (maxima of the vector potential), leading to interference patterns in the real-space density as the fast rescattered electrons overtake the slow direct electrons. Even though the pulse is several optical cycles long and the Keldish parameter $\gamma_{\mathrm{eff}} = 2.1 > 1$ is at the border to the multi-photon regime, the density plot is dominated by the two central field maxima where charge is emitted. This double slit in time leads to interferences in energy (multi-photon peaks), which are visible in the position-space density as fringes with increasing slope (examples marked with grey tilted arrows).

Inside the tip ($z \lesssim 0$), the induced density fluctuations act to screen the external laser field. Notably, they are several orders of magnitude larger than the emitted part of the density, but are still small in comparison to the bulk electron density $n_0 = 1.88 \times 10^{-2}$ a.u. The conduction band electrons can thus adjust adiabatically to the slowly varying perturbation of the external field (laser period $\approx$ 100 a.u.). The electron gas responds similar to a harmonic oscillator driven at frequencies far below its resonance since the photon energy of $\hbar \omega = 1.55$ eV is much smaller than the plasmon energy of $\hbar \omega_p = \sqrt{ 4 \pi n_0 } = 13.2$ eV. Fine scale Friedel oscillations (wavelength $ \sim \lambda_F / 2 = 3.8$ a.u.) are visible near the surface, similar to the response to a static field. Screening is complete within a few Friedel oscillations ($\lesssim 10$ a.u.).

\section{Dependence on exchange-correlation potential} \label{sec:vxc}
\begin{figure}[h] 
\centerline{\includegraphics[width=0.9\textwidth]{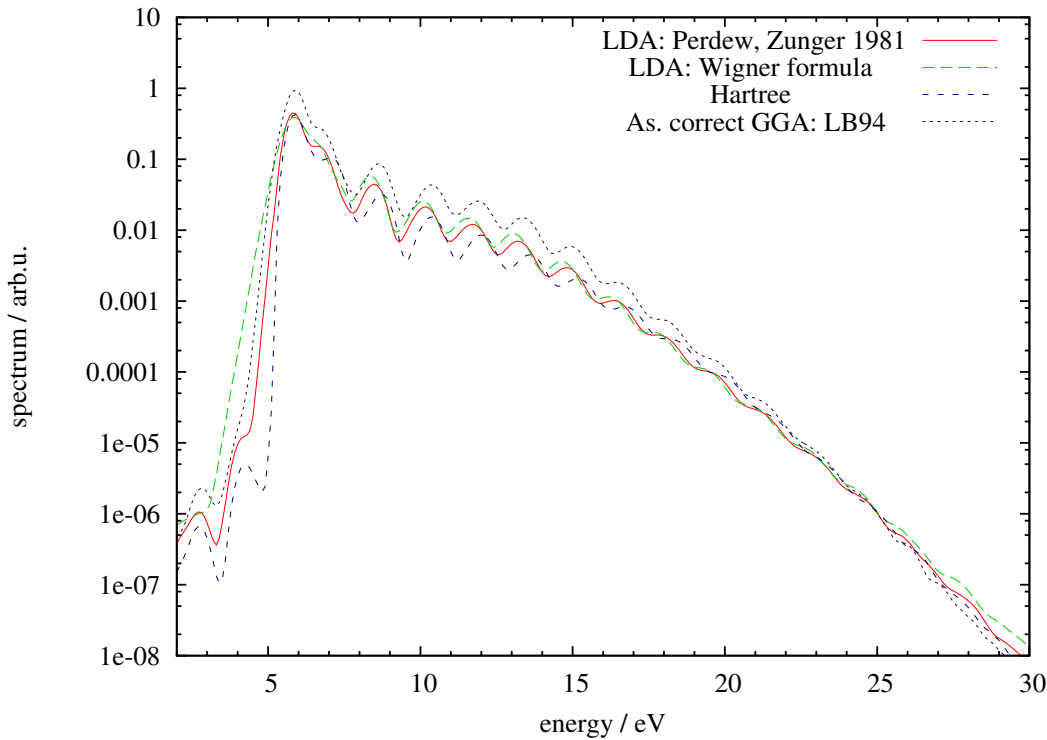}}
\caption{\label{fig:xcpot}
Calculated electron spectra for different exchange-correlation potentials. Local density approximation (LDA) using  the Perdew-Zunger parametrization \cite{Perdew1981Selfinteraction} (red solid line), LDA using the Wigner interpolation formula (green wide dashed line), exchange-correlation switched off  (Hartree approximation, blue short dashed line), asymptotically correct generalized gradient approximation (GGA) by van Leeuwen and Barends \cite{vanLeeuwen1994Exchangecorrelation} (purple dotted line).
}
\end{figure}

As an illustration of the theoretical error bar and the robustness of our simulation to details of the model  we show the dependence of our results on the exchange-correlation potential in fig.\ \ref{fig:xcpot}. Here, the enhanced laser field amplitude is $F_{\mathrm{eff}} = 0.021$ a.u.\ and the remaining parameters are taken as in section \ref{sec:samplesim}. We compare two different parametrizations of the local density approximation (LDA), the original Wigner formula (green long dashed line) and the improved parametrization by Perdew and Zunger (PZ81 \cite{Perdew1981Selfinteraction}, red solid line), the asymptotically correct generalized gradient approximation (GGA) by van Leeuwen and Baerends (LB94 \cite{vanLeeuwen1994Exchangecorrelation}, purple dotted line) and the Hartree approximation, i.e.\ switching off $V_{xc}$ (blue short dashed line). Remarkably, the main features of the spectra are well reproduced with all four XC potentials: the slope of the direct part, the height of the rescattered part relative to the direct part, and the position and slope of the cut-off are virtually indistinguishable. Slight deviations are found in phase-sensitive details of the spectra such as the visibility and position of the peaks, likely owing to details of the time-dependent mean field near the surface. The fact that the Hartree potential leads to reasonable results underscores that the time-dependent dynamics are strongly dominated by the external laser field and its time-dependent interaction with the electron density via the classical electrostatic interaction (Poisson equation).

\section{Dependence on atomic core potential} \label{sec:surfpot}
Backscattering happens with a high probability when the impinging electron wave function experiences a steep potential gradient. In strong field studies with atomic systems, this potential gradient is provided by the Coulomb potential of the atomic nuclei. On the other hand, our jellium description presented above smears out the potentials of the atomic nuclei to a background potential, resulting in a smooth single-particle potential with a small reflection coefficient. It is therefore of interest to compare simulated spectra with and without the additional potential mimicking an atomic core (eq.\ \ref{eq:vatom}). 

\begin{figure}[h]
\centerline{\includegraphics[width=0.9\textwidth]{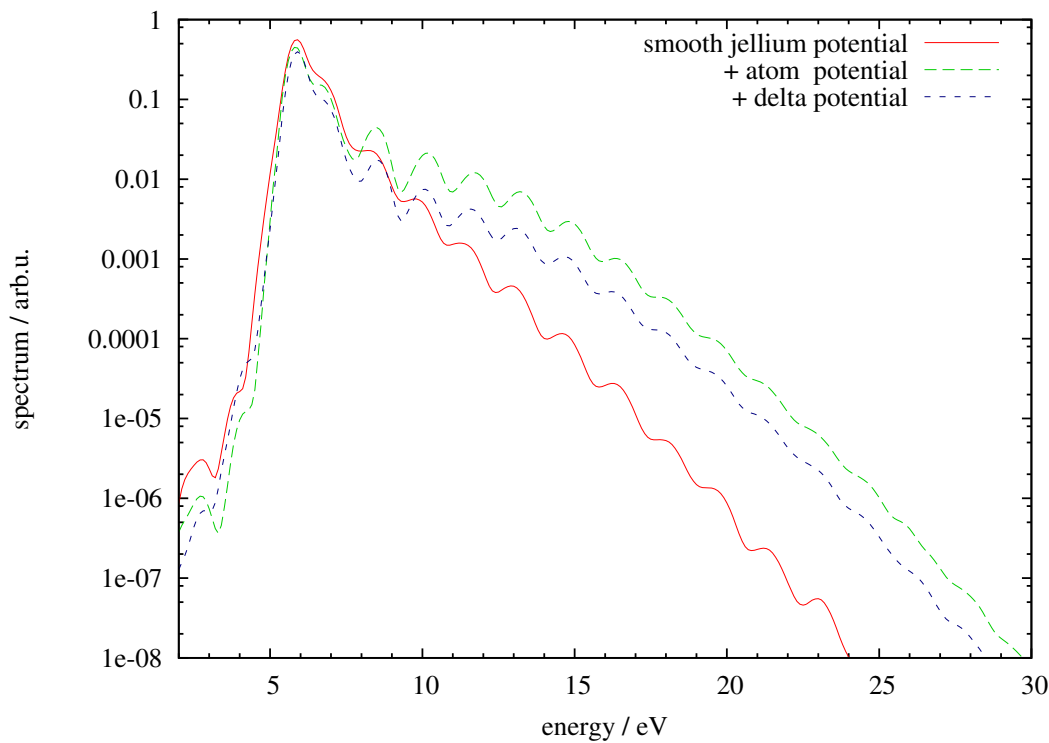}}
\caption{ \label{fig:surfpot}
Calculated electron spectra for a jellium potential (red solid line), jellium with added screened smoothed Coulomb potential (green wide dashed line), and jellium with added delta potential (blue short dashed line). The added potentials boost the rescattering probability to levels necessary for the observation of the plateau. Apart from this effect, the different model potentials do not change the calculated spectra. 
}
\end{figure}

Fig.~\ref{fig:surfpot} shows such a comparison of calculated spectra for an enhanced laser amplitude of $F_{\mathrm{eff}} = 0.021$ a.u. The low-energy parts of the spectra (first two peaks) look very similar but striking differences appear in the high energy parts. While a plateau structure is observed for a screened Coulomb potential (green wide dashed line) as well as a simple delta-like potential (blue short dashed line), a jellium potential alone (red solid line) does not show a plateau because the reflection coefficient is so low that it is covered by the direct part. We therefore infer that backscattering is indeed induced by atomic cores near the surface. 

We note that in fig.~\ref{fig:surfpot} the essential features of the spectrum (decay, plateau, and cutoff) are qualitatively reproduced equally well by a screened Coulomb potential or by a delta function. The only difference lies in the relative height of the plateau which reflects the magnitude of the reflection coefficient for recolliding electrons. Details of the scattering potential, encoded in the scattering phase, cancel out in the interferences within the rescattered wave packets \cite{Kruger2012Interaction} and leave no signature in the electron spectra under the present conditions.

\section{Conclusions}
We have presented a theoretical study of electron emission from nano-scale metal tips by near-infrared few cycle laser pulses of moderate intensity. We have probed the sensitivity of the simulation to the employed approximation to the exchange-correlation potential demonstrating that exchange and correlation effects play only a small role. The dependence of the simulated electron spectra on the atomic core pseudo-potential has been investigated corroborating that rescattering is induced by the atomic cores of the atoms in the first layer. 

\section*{Acknowledgments}
This work was supported by the Austrian Science Foundation FWF under Proj.\ Nos.\ SFB-041 ViCoM and P21141-N16. G.W.\ thanks the International Max Planck Research School of Advanced Photon Science for financial support. The authors would like to thank M.\ Schenk, M.\ Kr\"uger and P.\ Hommelhoff for discussions.

\section*{References}

\bibliography{georgwachter_nourl}

\end{document}